\documentclass[aps,twocolumn,groupedaddress,showpacs,floats,amssymb]{revtex4}
\usepackage[dvips]{graphicx}
\usepackage{bm}
 
\begin{document}

\noindent 
{\bf Comment on ``Phase Diagram of a Disordered Boson 
Hubbard Model in Two Dimensions''}

\bigskip

In a recent Letter \cite{LeeChaKim} (see also \cite{Kisker}) 
the authors presented numerical evidence supporting an idea of a direct 
transition between the superfluid (SF) and  Mott insulating (MI) 
phases in the disordered Bosonic system, and even studied non-trivial 
properties of the multicritical line where SF, MI and the Bose Glass (BG) phases 
meet. The results were obtained from Monte Carlo simulations 
of the (2+1)-dimensional classical loop-current model \cite{Wallin94}
with the lattice action 
\begin{equation}
S =  {1 \over 2K} \sum_{{\bf r} \tau }^{\nabla \cdot \vec{J}=0} 
\bigg[ \vec{J}^2({\bf r},\tau ) -
 2 (\mu +v({\bf r}) ) \vec{J}_{\tau }({\bf r},\tau ) \bigg] \;.
\label{model}
\end{equation}
where  ${\bf r}, \tau$ are spatial and imaginary time coordinates,
and $\vec{J}({\bf r},\tau )$ are integer current vectors with zero
divergence. The spatial disorder potential $v({\bf r})$ is uniformly
distributed on the interval $(-\Delta , \Delta )$.
 
Here we prove that all of the above mentioned conclusions are incorrect
and originate from doing simulations for too small system sizes 
(the maximum system size considered in \cite{LeeChaKim} was 
$L_x  \times L_y  \times L_{\tau } = 14 \times 14 \times 20$),
and ignoring the rigorous theorem of \cite{Fisher89,FM96} 
saying that for disorder
$\Delta$ larger than the half-width of the energy gap in the 
ideal Mott insulator (we denote it as $E_g$) the system state is 
compressible, i.e. it is BG. Indeed, in the infinite system one can 
always find arbitrary large regions with the chemical potential 
being nearly homogeneously shifted downwards or upwards by $\Delta$. There is no
energy gap then for the particle transfer between such regions, and they 
can be doped with particles/holes.
[The conjecture is that the MI-BG transition is exactly at the 
upper boundary of the theorem, $\Delta = E_g(K)$.]
We note, that the theorem is based on rare statistical fluctuations, 
and for $\Delta \to 0$  the distance between regions contributing 
to non-zero but {\it exponentially} small compressibility is 
diverging {\it exponentially}.

In Fig.~1 we plot our data for $E_g(K)$ \cite{ourcondmat2003}
along with the critical values of disorder for the superfluid-insulator transition.
Since $\Delta $ is always {\it larger } than $E_g(K_c)$, the
transition is always from SF to the compressible insulating phase, or BG. 
Accordingly, the multicritical line where SF, MI, and BG phases meet 
does not exists for non-zero $\Delta$.
  
The diverging distance between rare statistical realizations of disorder
is the major problem in interpreting Monte Carlo data for small
system sizes. Recently developed Worm algorithms \cite{ourcondmat2003,AS} 
allow simulations of system sizes as large as $160 \times 160 \times 1000$, 
but for $\Delta =0.4$ even this is barely enough to resolve 
very small but finite compressibility at the critical point \cite{ourcondmat2003}. 
Clearly, for smaller system sizes and smaller values of disorder the 
short length-scale behavior will mimic a direct SF-MI transition observed in 
Refs.~\cite{Kisker,LeeChaKim}. Similar arguments were used previously 
in the criticism of the direct SF-MI transition in one dimensional 
systems \cite{ourcomment98}.

 
\begin{figure}[tp]
\includegraphics[width=6.5cm]{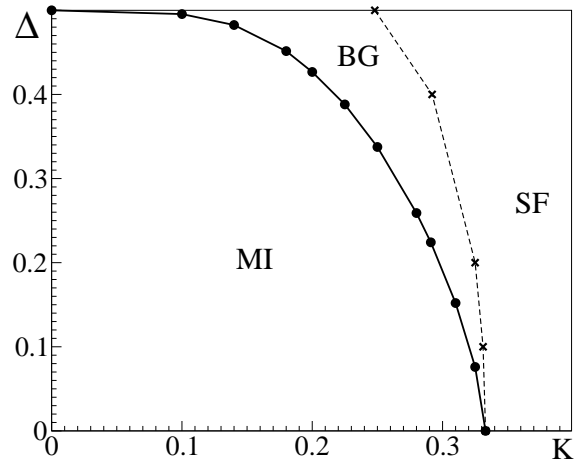}
\vspace*{-1cm}
\caption{ The phase diagram of the disordered,
commensurate loop-current model (\ref{model}). 
All error bars are of order $10^{-3}$ and smaller than point sizes;
data points for the SF-BG line were taken from: 
$\Delta =0.4$ \cite{Kisker,LeeChaKim,ourcondmat2003},
$\Delta =0.2$ \cite{Kisker}
$\Delta =0.1,~~0.5$ [this work].} 
\label{FigComment}
\end{figure}

 
This work was supported by the National Science Foundation under
Grant DMR-0071767.

\bigskip
 
\noindent  Nikolay Prokof'ev and Boris Svistunov              \\ \indent 
{\small    Department of Physics, }                           \\ \indent 
{\small    University of Massachusetts, Amherst, MA 01003, }  \\ \indent 
{\small    Russian Research Center ``Kurchatov Institute'', } \\ \indent 
{\small    123182 Moscow, Russia}

\bigskip
 


\bigskip
\def\refname{}


\begin{thebibliography}{99}

\bibitem{LeeChaKim} J.-W. Lee, M.-C. Cha, and D. Kim,
                    Phys. Rev. Lett. {\bf 87}, 247006 (2001);
                    {\it ibid}, {\bf 88}, 049901 (2002) 
                    
\bibitem{Kisker} J. Kisker and H. Rieger,
                 Phys. Rev. B {\bf 55}, R11981 (1997).
                 
\bibitem{Wallin94}  M. Wallin, {\it et al.},
                    Phys. Rev. B {\bf 49}, 12115 (1994).
      
\bibitem{Fisher89} M.P.A. Fisher, {\it et al.},
                   Phys. Rev. B, {\bf 40}, 546 (1989);
 
\bibitem{FM96} J.K. Freericks and H. Monien,
               Phys. Rev. B {\bf 53}, 2691 (1996).

\bibitem{ourcondmat2003} N.V. Prokof'ev and  B.V. Svistunov,  
                         cond-mat/0301205.
                         
\bibitem{AS} F.~Alet and E.S.~S{\o}rensen,
             Phys. Rev. E {\bf 67}, 015701 (2003);
             {\it ibid}, {\bf 68}, 026702 (2003).
                                         
               
\bibitem{ourcomment98}  N.V. Prokof'ev and B.V. Svistunov, 
                        Phys. Rev. Lett. {\bf 80}, 4355 (1998);
  
\end{thebibliography}
\end{document}